\documentclass[twocolumn,showpacs,preprintnumbers,amsmath,amssymb]{revtex4}
\usepackage{graphicx}
\usepackage{dcolumn}
\usepackage{bm}
\usepackage{subfigure}
\usepackage{color}

\begin{document}

\title{Perturbative and nonperturbative correlations in double parton distributions}

\author{A.M.~Snigirev}
\affiliation{Skobeltsyn Institute of Nuclear Physics, Lomonosov Moscow State University, 119991, Moscow, Russia}

\author{N.A.~Snigireva}
\affiliation{School of Mathematical Sciences, University College Dublin, Belfield, Dublin 4, Ireland}

\author{G.M.~Zinovjev}
\affiliation{Bogolyubov Institute for Theoretical Physics, National Academy of Sciences of Ukraine, 03680 Kiev-143, Ukraine}

\date{\today}
\begin{abstract}
We argue that the perturbative QCD correlations contribute dominantly to the double parton distributions as compared to the nonperturbative ones in the limit of
sufficiently large hard scales and for not parametrically small longitudinal momentum fractions.   
\end{abstract}


\maketitle
\section{\label{sec1}Introduction}

The process of multiple parton interactions (MPI) becomes the important factor in the data analysis due to its substantial contribution to the background~\cite{DelFabbro:2002pw,Hussein:2006xr,Bandurin:2010gn} in studying the Higgs production and other interesting phenomena in the LHC experiments. The splash of research activity around MPI in recent years~\cite{Bartalini:2011jp,Abramowicz:2013iva} has been stimulated by the experimental evidence for double parton scattering (DPS) in the processes  producing two independently identified hard particles. Such processes have been observed in $pp$ and $p\bar p$ collisions for the final states containing four jets, $\gamma+3$ jets, and $W+2$ jets by AFS~\cite{AFS}, UA2~\cite{UA2}, CDF~\cite{cdf4jets,cdf}, D0~\cite{D0,D01,D02}, ATLAS~\cite{atlas}, and CMS~\cite{cms} Collaborations. The data on double $J/\psi$ production~\cite{Aaij}, and single $J/\psi$ production as a function of the event multiplicity~\cite{Abelev}
can also be successfully interpreted using  DPS~\cite{kom, Baranov:2011ch, Novoselov, Baranov:2012re} and  MPI models, respectively.

The theoretical treatment and the phenomenological analysis of DPS dates back to the early days of the parton model, and the DPS concept was subsequently
extended to the perturbative QCD (see, for instance, the reviews~\cite{Bartalini:2011jp,Abramowicz:2013iva} with many references to the previous works therein).
The role of the DPS mechanism in application to the proton-nucleus and nucleus-nucleus collisions was recently investigated
in Refs.~\cite{Strikman:2001gz,Treleani:2012zi,d'Enterria:2012qx,d'Enterria:2013ck,Blok:2012jr,Calucci:2013pza,Blok:2014ora,d'Enterria:2014bga}.  
Meanwhile the MPI phenomenology rests upon the physically intuitive considerations and involves substantial simplifying assumptions, therefore it seems
desirable to unite and strengthen theoretical efforts in order to achieve a better description of MPI, and in particular DPS, which may very likely be the most
significant multiple scattering mode at the LHC.

The initial state of DPS is defined by the double parton distribution functions (dPDFs) which quantify the joint distribution of two partons in a hadron, depending on their quantum numbers, longitudinal momentum fractions, and the relative transverse distance between them. The starting cross section formula for DPS is somewhat similar to that commonly used for the single parton scattering (SPS). It was derived by making use of the light-cone variables and the same approximations as those applied to the processes with a single hard scattering. The inclusive DPS cross section (in the momentum representation) for the hadron-hadron collision with two hard parton subprocesses $A$ and $B$ may be written in the factorized form as
\begin{eqnarray}
\label{hardAB_p}
\sigma_{\rm DPS}^{(A,B)} =& & \frac{m}{2} \sum \limits_{i,j,k,l} \int \Gamma_{ij}(x_1, x_2; {\bf q}; Q^2_1, Q^2_2)\nonumber\\
& &\times \hat{\sigma}^A_{ik}(x_1, x_1^{'}) 
\hat{\sigma}^B_{jl}(x_2, x_2^{'}) \Gamma_{kl}(x_1^{'}, x_2^{'}; {\bf -q}; Q^2_1, Q^2_2)\nonumber\\ 
& & \times dx_1 dx_2 dx_1^{'} dx_2^{'} \frac{d^2q}{(2\pi)^2},
\end{eqnarray}
where $\hat{\sigma}^A_{ik}$ and $\hat{\sigma}^B_{jl}$ are the parton-level subprocess cross sections, $\Gamma_{ij}(x_1, x_2; {\bf q} ; Q^2_1, Q^2_2)$ are the generalized dPDFs, depending on the longitudinal momentum fractions $x_1$ and $x_2$ of the two partons $i$ and $j$ undergoing the hard processes $A$ and $B$ at the scales $Q_1$ and $Q_2$. The combinatorial factor $m/2$ accounts for the indistinguishable ($m=1$) and  distinguishable ($m=2$) final states. The dPDFs in the momentum representation depend on the transverse momentum ${\bf q}$ which is equal to the difference of the momenta of partons from the wave function of the colliding hadrons in the amplitude and the amplitude conjugated. Such a dependence arises because the difference of parton transverse momenta within the parton pair is not conserved. This transverse momentum {\bf q} is the Fourier conjugated variable of the parton pair transverse separation which is used in the mixed (momentum and coordinate) representation.

The dPDFs and the corresponding evolution equations are well-known~\cite{Kirschner:1979im,Shelest:1982dg,snig03,snig04} only for ${\bf q}=0$ (in other words, integrated over the parton pair transverse separation) in the collinear approximation. In this approximation the two-parton distribution functions, 
$$\Gamma_{ij}(x_1, x_2; {\bf q}=0; Q^2, Q^2)=D^{ij}_h(x_1, x_2; Q^2, Q^2),$$
with the two hard scales are set equal and satisfy the generalized Dokshitzer-Gribov-Lipatov-Altarelli-Parisi (DGLAP) evolution equations, derived initially in Refs.~\cite{Kirschner:1979im,Shelest:1982dg}. The subsequent extension to two different hard scales was done in Ref.~\cite{Ceccopieri:2010kg}, whereas the single parton distributions satisfy the widely known DGLAP equations~\cite{gribov,lipatov,dokshitzer,altarelli}. The functions in question have a specific interpretation in the leading logarithm approximation of perturbative QCD. They are the inclusive probabilities which allow one to find two bare partons of types $i$ and $j$ with the given longitudinal momentum fractions $x_1$ and $x_2$ in a hadron $h$.

These well-known collinear distributions were a starting point to derive the revised formula for inclusive cross section of the 
DPS process without the simplifying {\em additional} factorization assumption (which, in general, is inconsistent with the QCD evolution) suggested in Ref.~\cite{Ryskin:2011kk}. The new formula contains three contributing terms: (i) the "traditional" factorization component, (ii) the single and (iii) the double perturbative splitting graphs induced by the QCD evolution. Later on,  similar results were also obtained in Ref.~\cite{Blok:2011bu}, with an emphasis on the differential cross sections, and were partly corroborated in Refs.~\cite{Gaunt:2012wv,Gaunt:2012dd}, albeit with some distinctions attributed mainly to the terminology. It was found in Refs.~\cite{Ryskin:2011kk, Ryskin:2012qx} that the so-called single and double perturbative splitting graphs can significantly contribute to the inclusive cross section for the DPS process as compared to the "traditional" factorization component. It makes sense to mention here the discussion~\cite{stir,Blok:2011bu,Gaunt:2012wv,Ryskin:2012qx,Manohar,Gaunt:2012dd,Diehl:2011tt} concerning the double perturbative splitting graphs. Formally, this contribution within the collinear approach in the region of not too small $x$ should be considered as a result of the interaction of {\it one} parton pair with the $2\to 4$ hard subprocess~\cite{stir,Blok:2011bu,Gaunt:2012wv,Ryskin:2012qx,Manohar,Gaunt:2012dd}, since the dominant contribution to the phase space integral comes from a large $q^2\sim \min(Q_1^2,Q_2^2)$. However, as it was argued in Refs.~\cite{Ryskin:2011kk,Ryskin:2012qx}, the contribution under discussion may be validly included in the DPS cross section for appropriately low longitudinal momentum fractions. 

In any case the numerical evaluation of single and double perturbative splitting graphs contributing to the DPS cross section is desirable at the LHC kinematics, where the large available values of $Q_1$ and $Q_2$ (in comparison with the characteristic QCD small reference scale $\mu$), $\ln{(1/x_1)}$ and $\ln{(1/x_2)}$ (in comparison with 1) can provide the configurations with the Balitsky-Fadin-Kuraev-Lipatov (BFKL)~\cite{ryskin,bfkl,bfkl2,bal} or the DGLAP evolution in the ladders before and after splitting, which depend on the features of the considered processes. Such estimates for the contribution of single perturbative splitting graphs to the DPS cross section were recently done in Ref.~\cite{Blok:2013bpa} (the three-parton interactions in the authors' terminology), and it was pointed out that the relative contribution of the evolution effects increases with increasing hard scales and  may resolve the long standing puzzle: why the observed effective DPS cross section is underestimated by a factor of two in the independent parton approximation (with regard to the "traditional" factorization component only)? 

The main purpose of the present paper is to demonstrate analytically that the perturbative QCD correlations contribute dominantly to
dPDFs as compared with the nonperturbative ones in the limit of sufficiently large hard scales and at parametrically not small longitudinal momentum fractions --- the conclusion definitely supporting the numerical results of Ref.~\cite{Blok:2013bpa}. The paper is organized as follows. In Sec.~\ref{sec2} we discuss the two-parton distribution functions resultant from the perturbative QCD theory and introduce the appropriate definitions and designations. The asymptotic behavior of dPDFs is discussed in Sec.~\ref{sec3}, and in Sec.~\ref{sec4} we summarize the obtained results.

\section{\label{sec2}Double parton distributions in the leading logarithm approximation}

The analysis performed in Refs.~\cite{gribov, lipatov, dokshitzer} for the hard processes (deep-inelastic electron-proton scattering and electron-positron annihilation into hadrons) in vector, pseudoscalar and QCD field theories has provided a powerful tool --- the leading logarithm approximation in terms of the parton model with a variable cutoff parameter for the transverse momenta. The dependence of multiparton distribution and fragmentation functions on the value of this cutoff parameter is determined by the evolution equations. The most transparent method for deriving such equations in any renormalizable quantum field theory was formulated in Ref.~\cite{lipatov}. The value of hard scale (most frequently, the transfer momentum squared $Q^2$), or its logarithm $\xi=\ln(Q^2/\mu^2)$, or double logarithm (which takes into account explicitly the behavior of the effective coupling constant in the leading logarithm approximation) is treated as the evolution variable
\begin{eqnarray}
t & = &\frac{1}{2\pi \beta} \ln \Bigg[1 + \frac{g^2(\mu^2)}{4\pi}\beta
\ln\Bigg(\frac{Q^2}{\mu^2}\Bigg)\Bigg] \nonumber \\
& = &\frac{1}{2\pi \beta}\ln\Bigg
[\frac{\ln(\frac{Q^2}{\Lambda^2_{\rm QCD} })}
{\ln(\frac{\mu^2}{\Lambda^2_{\rm QCD}})}\Bigg].
\end{eqnarray}
Here $\beta = (33-2n_f)/12\pi$ ${\rm {in~ QCD}}$, $g(\mu^2)$ is the running coupling constant at some characteristic scale $\mu^2$ above which the perturbative theory is applicable, $n_f$ is the number of active flavors and $\Lambda_{\rm QCD}$ is the QCD dimensional parameter.

The DGLAP evolution equations~\cite{gribov,lipatov, dokshitzer, altarelli} assume the simplest form if we use the natural dimensionless evolution variable $t$; that is, 
\begin{equation}
\label{e1singl}
 \frac{dD_i^j(x,t)}{dt} = 
\sum\limits_{j{'}} \int \limits_x^1
\frac{dx{'}}{x{'}}D_i^{j{'}}(x{'},t)P_{j{'}\to j}\Bigg(\frac{x}{x{'}}\Bigg).
\end{equation}
These equations describe the evolution of single distributions $ D^j_i(x,t)$ of bare quarks and gluons within dressed partons (quarks and gluons, $i,j = q/g$) in response to the change of evolution variable $t$. The kernels, $P$, of these equations in the Lipatov's method already include a regularization at $x \rightarrow x{'}$ unlike the regularization in Ref.~\cite{altarelli} where it was introduced {\it ad hoc} by requiring the momentum conservation. Moreover, this method makes it possible to derive evolution equations for multiparton distribution (and fragmentation) functions as well.

These equations were derived for the first time~\cite{Kirschner:1979im, Shelest:1982dg} in the following form
\begin{eqnarray}
\label{edouble}
& &\frac{dD_i^{j_1j_2}(x_1,x_2,t)}{dt} \\
& &=\sum\limits_{j_1{'}}
\int\limits_{x_1}^{1-x_2}\frac{dx_1{'}}{x_1{'}}D_i^{j_1{'}j_2}(x_1{'},x_2,t)
P_{j_1{'}
\to j_1} \Bigg(\frac{x_1}{x_1{'}}\Bigg) \nonumber\\
& &+ \sum\limits_{j_2{'}}\int\limits_{x_2}^{1-x_1}
\frac{dx_2{'}}{x_2{'}}D_i^{j_1j_2{'}}(x_1,x_2{'},t)P_{j_2{'} \to j_2}
\Bigg(\frac{x_2}{x_2{'}}\Bigg) \nonumber\\
& &+ \sum\limits_{j{'}}D_i^{j{'}}(x_1+x_2,t) \frac{1}{x_1+x_2}P_{j{'} \to
j_1j_2}\Bigg(\frac{x_1}{x_1+x_2}\Bigg).\nonumber
\end{eqnarray}
Here, the splitting kernel
\begin{equation}
\frac{1}{x_1+x_2} P_{j{'} \to
j_1j_2}(\frac{x_1}{x_1+x_2}),
\end{equation}
which appears in the nonhomogeneous part of the equations, does not include the $\delta$-function regularizing term. The equations describe the dPDFs evolution of bare quarks and gluons in dressed partons (quarks and gluons) as a function of the evolution variable $t$ --- that is, for the case where the scales of the two hard processes are commensurate ($Q_1^2\simeq Q_2^2$) so that there is no another large logarithm $|\ln(Q_1^2/Q_2^2)|$. Thus we remain within the leading logarithm approximation, in which, we recall, the variable $\xi$ is specified apart from a constant because of the ambiguous choice of the reference scale $\mu^2$.

It is easy to make certain by direct substitution that the solutions of Eq.~(\ref{edouble}) can be written as a convolution of single distributions~\cite{Kirschner:1979im, Shelest:1982dg}
\begin{eqnarray}
\label{solution}
& & D_i^{j_1j_2}(x_1,x_2,t) \\
& &= \sum\limits_{j{'}j_1{'}j_2{'}} \int\limits_{0}^{t}dt{'}
\int\limits_{x_1}^{1-x_2}\frac{dz_1}{z_1}
\int\limits_{x_2}^{1-z_1}\frac{dz_2}{z_2}~
D_i^{j{'}}(z_1+z_2,t{'})\frac{1}{z_1+z_2} \nonumber\\
& & \times P_{j{'} \to
j_1{'}j_2{'}}\Bigg(\frac{z_1}{z_1+z_2}\Bigg) D_{j_1{'}}^{j_1}(\frac{x_1}{z_1},t-t{'}) 
D_{j_2{'}}^{j_2}(\frac{x_2}{z_2},t-t{'}).\nonumber
\end{eqnarray}
\noindent 
This expression coincides with the jet calculus rules~\cite{konishi0, konishi} proposed originally for the fragmentation functions, and is a generalization of the Gribov-Lipatov relation installed for single functions~\cite{gribov, lipatov, dokshitzer} (the distribution of bare partons inside a dressed constituent is identical to the distribution of dressed constituents in the fragmentation of a bare parton in the leading logarithm approximation). Nevertheless, the direct numerical integration of Eq.~(\ref{edouble}) seems to be more effective~\cite{Gaunt:2009re} than the explicit solutions~(\ref{solution}) for the phenomenological treatment because of the singular $\delta$-like initial conditions of single distributions $D_i^j(x,t)$ (the Green's functions) figuring in these solutions. The solutions presented in (\ref{solution}) show that in the leading logarithm approximation dPDFs are strongly correlated at the parton level; that is,
\begin{eqnarray}
\label{nonfact}
D_i^{j_1j_2}(x_1,x_2,t) \neq D_{i}^{j_1}(x_1,t) 
D_{i}^{j_2}(x_2,t).
\end{eqnarray}

The distributions of bare quarks and gluons in a {\it hadron} are more interesting for phenomenological applications. Apparently,
such distributions may be obtained within the well-known approach based upon factorization of soft and hard stages (so-called physics of short and long distances~\cite{collins}). As a result, Eqs.~(\ref{e1singl}) and (\ref{edouble}) will describe the evolution with $t(Q^2)$ of parton distributions in a hadron, if we only substitute the index $i$ by index $h$. However, the initial conditions for new equations at $t=0(Q^2=\mu^2)$ are unknown {\it  a priori} and should be introduced phenomenologically --- e.g. extracted from the experiment or taken from some model dealing with the physics of long distances [at the parton level, 
$D_{i}^{j}(x,t=0)= \delta_{ij} \delta(x-1)$ and $D_i^{j_1j_2}(x_1,x_2,t=0)=0$].
The solutions of the generalized DGLAP evolution equations with the given initial conditions may be written, as before, in form of a convolution of single distributions:
\begin{eqnarray}
\label{solution1}
D_h^{j_1j_2}(x_1,x_2,t) = D_{h1}^{j_1j_2}(x_1,x_2,t)+ D_{h({\rm QCD})}^{j_1j_2}(x_1,x_2,t),
\end{eqnarray}
where
\begin{eqnarray}
\label{solnonQCD}
 D_{h1}^{j_1j_2}(x_1,x_2,t)&=&\sum\limits_{j_1{'}j_2{'}} 
\int\limits_{x_1}^{1-x_2}\frac{dz_1}{z_1}
\int\limits_{x_2}^{1-z_1}\frac{dz_2}{z_2}~
D_h^{j_1{'}j_2{'}}(z_1,z_2,0) \nonumber\\
& & \times D_{j_1{'}}^{j_1}(\frac{x_1}{z_1},t) 
D_{j_2{'}}^{j_2}(\frac{x_2}{z_2},t) ~,
\end{eqnarray}
and
\begin{eqnarray}
\label{solQCD}
& & D_{h({\rm QCD})}^{j_1j_2}(x_1,x_2,t) \\
& &= \sum\limits_{j{'}j_1{'}j_2{'}} \int\limits_{0}^{t}dt{'}
\int\limits_{x_1}^{1-x_2}\frac{dz_1}{z_1}
\int\limits_{x_2}^{1-z_1}\frac{dz_2}{z_2}~
D_h^{j{'}}(z_1+z_2,t{'})\frac{1}{z_1+z_2} \nonumber\\
& & \times P_{j{'} \to
j_1{'}j_2{'}}\Bigg(\frac{z_1}{z_1+z_2}\Bigg) D_{j_1{'}}^{j_1}(\frac{x_1}{z_1},t-t{'}) 
D_{j_2{'}}^{j_2}(\frac{x_2}{z_2},t-t{'})\nonumber
\end{eqnarray}
are the dynamically correlated distributions originating from the perturbative QCD (compare (\ref{solution}) and (\ref{solQCD})).

The first term of the generalized solutions is solutions of homogeneous evolution equations (independent evolution of two branches), where  the input two-parton distributions are generally not known at the low scale limit $\mu (t=0)$. For these non-perturbative two-parton functions at low $z_1,z_2$ one may assume the factorization $D_h^{j_1{'}j_2{'}}(z_1,z_2,0) \simeq D_h^{j_1{'}}(z_1,0)D_h^{j_2{'}}(z_2,0)$ neglecting the constraints imposed by the momentum conservation ($z_1+z_2<1$). This leads to
\begin{eqnarray} 
\label{DxD_Q}
 D^{ij}_{h1}(x_1, x_2,t)
 \simeq D^i_h (x_1,t) D^j_h (x_2,t)
\end{eqnarray}
and justifies partly the factorization hypothesis for dPDFs usually applied in practical calculations.

Surely, it is interesting to know the magnitude of induced correlations with respect to the factorization component. Numerically, the contribution of these evolution-induced correlations was estimated in Ref.~\cite{snig04}. The required for such calculations initial data for single parton distributions $D_h^i(x,0)$ were specified at the scale $Q_0=\mu= 1.3$ GeV in accordance with the parametrization established in the CTEQ Collaboration~\cite{cteq}. The ratio of the gluon-gluon correlations in the proton resulting from evolution, to the factorized component
\begin{eqnarray}
\label{ratio}
R(x,t)=
\frac{D_{p({\rm QCD,corr.})}^{gg}(x_1,x_2,t)}{ D_p^{g}(x_1,t)D_p^{g}(x_2,t)(1-x_1-x_2)^{2}} \Big|_{x_1=x_2=x}
\end{eqnarray}
was calculated and it appeared that, at the hard scale of the CDF measurements ($Q \sim 5$ GeV) the ratio (\ref{ratio}) is nearly 10$\%$ and increases up to 30$\%$ at much higher scale ($Q \sim 100$ GeV) for the longitudinal momentum fractions $x \leq 0.1$ accessible in these measurements. The correlations may increase up to 90$\%$ for the finite longitudinal momentum fractions $x \sim 0.2 \div 0.4$, and become important for almost all $x$ with increasing $t$. 

Here it should be emphasized that the momentum conserving phase space factor $(1 - x_1 - x_2)^2$ is introduced in Eq.~(\ref{ratio}) instead of $(1 - x_1 - x_2)$ usually used. The reason is quite obvious again --- this factor is introduced, generally speaking, "by hand" in order to "secure" the momentum conservation, i.e. in order that the product of two single distributions would vanish at
$x_1 + x_2 = 1.$ However, the generalized QCD evolution equations demand higher power of $(1 - x_1 - x_2)$ at $(x_1 + x_2) \rightarrow 1$: the phase space integrals in Eqs. (\ref{solution}), (\ref{solnonQCD}) and (\ref{solQCD}) only give
$$
\int\limits_{x_1}^{1-x_2} dz_1
\int\limits_{x_2}^{1-z_1}dz_2~ =~(1 - x_1 - x_2)^2/2 .$$
In fact, this exponent should depend on $t$ (increase with increasing $t$) as it takes place for single distributions at $x \rightarrow 1$~\cite{dokshitzer}. The numerical calculations also support this conclusion:  the exponent of $(1 - x_1 - x_2)$ for the perturbative QCD gluon-gluon correlations is found to be larger than 2 and increasing with $t(Q)$. However, the introduced factor $(1 - x_1 - x_2)^2$ does not practically affect the ratio (\ref{ratio}) in the region of small $x_1$ and $x_2$, while just this region, in which multiple interactions can contribute to the cross section noticeably, is especially interesting from the experimental point of view.

The properties of dPDFs in hadrons were studied in more detail~\cite{Gaunt:2009re,Diehl:2014vaa} by integrating directly the evolution equations~(\ref{edouble}) (in Ref.~\cite{Diehl:2014vaa} only homogeneous evolution equations). This method seems to be more practicable since one does not need to deal with the singular Green functions (single parton level functions satisfying singular $\delta$-like conditions). These numerical estimations confirm also that the evolution effects are getting larger with increasing hard scales. 

Note that the particular solutions (\ref{solQCD}) of non-homogeneous equations contribute to the inclusive cross section of DPS with a larger weight (different effective cross section)~\cite{Ryskin:2011kk, Blok:2011bu, Ryskin:2012qx, Blok:2013bpa,Gaunt:2012dd,Cattaruzza:2005nu} as compared to the solutions (\ref{solnonQCD}) of homogeneous equations.  The latter solutions are usually approximated by a factorized form if the initial nonperturbative correlations are absent. These initial correlation conditions are {\it a priori} unknown yet not quite arbitrary as they obey the nontrivial sum rules~\cite{Gaunt:2009re,Ceccopieri:2014ufa} which are imposed upon the evolution equations. The problem of specifying the initial correlation conditions for the evolution equations, which would obey exactly these sum rules and have the correct asymptotic behavior near the kinematical boundaries, has been extensively studied in Refs.~\cite{Gaunt:2009re,Snigirev:2010ds,Chang:2012nw,Rinaldi:2013vpa,Golec-Biernat:2014bva,Ceccopieri:2014ufa}. Fortunately, the explicit form of evolution equation solutions allows us to answer the question:  which correlations (perturbative (\ref{solQCD}) or nonperturbative (\ref{solnonQCD})) are more significant at sufficiently large hard scale.

\section{\label{sec3}Asymptotic behavior}

Indeed, the evolution equations are explicitly solved by introducing the Mellin transformations
\begin{eqnarray}
\label{mellin}
 & M_h^{j}(n,t) = \int\limits_{0}^{1}dx x^n~D_{h}^{j}(x,t),\\
 &M_h^{j_1 j_2}(n_1,n_2,t) \nonumber\\
&= \int\limits_{0}^{1}dx_1 dx_2 \theta(1-x_1-x_2)x_1^{n_1}x_2^{n_2}D_{h}^{j_1 j_2}(x_1,x_2,t),
\end{eqnarray}
which lead to a system of ordinary linear differential equations of the first order:
\begin{equation}
\label{1.19}
dM_h^j (n,t)/dt = \sum\limits_{j{'}} M_h^{j{'}}(n,t) P_{j{'} \to j}(n), 
\end{equation}
\begin{eqnarray}
\label{1.17}
& &dM_h^{j_1j_2}(n_1,n_2,t)/dt \nonumber \\
& &=\sum\limits_{j_1{'}} M_h^{j_1{'}j_2} (n_1,n_2,t)
P_{j_1{'} \to j_1} (n_1) \nonumber\\
& &+ \sum\limits_{j_2{'}} M_h^{j_1j_2{'}}(n_1,n_2,t) P_{j_2{'} \to j_2}(n_2) 
\nonumber \\
& &+ \sum\limits_{j{'}} M_h^{j{'}} (n_1+n_2,t)P_{j{'} \to j_1j_2}(n_1,n_2), 
\end{eqnarray}
where the kernels,
\begin{eqnarray}
\label{1.21}
& P_{j' \to j}(n)  =  \int\limits_0^1 x^n P_{j' \to j} (x)dx, \\
 & P_{j' \to j_1 j_2}(n_1,n_2)  = \int\limits_0^1 x^{n_1} (1-x)^{n_2} P_{j' \to
j_1 j_2} (x) dx,   
\end{eqnarray}
are well-known and can be found in the explicit form --- as, for instance, in Refs.~\cite{dokshitzer, konishi,dok2}.

In order to obtain the distributions in $x$ representation, an inverse Mellin transformation should be performed
\begin{eqnarray}
\label{mellin in}
x D_h^{j}(x,t) =~\int\frac {dn}{2\pi i} x^{-n}~M_{h}^{j}(n,t),
\end{eqnarray}
\begin{eqnarray}
\label{mellin in2}
& &x_1 x_2 D_h^{j_1 j_2}(x_1,x_2,t)\nonumber \\
& &=~\int\frac {dn_1}{2\pi i} x_1^{-n_1}\int\frac {dn_2}{2\pi i} x_2^{-n_2}~M_{h}^{j_1 j_2}(n_1,n_2,t),
\end{eqnarray}
where the integration runs along the imaginary axis to the right hand side from all $n$ singularities. In the general case this can be done only numerically. However, the asymptotic behavior can be estimated in some interesting and particularly simple limits using the same technique as above.

Note that the exact solutions for single distributions in the moment representation can be written symbolically in a matrix form:
\begin{eqnarray}
\label{solution n}
M_i^{j}(n,t) =[\exp{P(n) t}]_i^j,
\end{eqnarray}
and the solutions of the generalized DGLAP evolution equations with the given initial conditions may be written again as a convolution of single distributions; in the moment representation, they read
\begin{eqnarray}
\label{1.37}
& & M_h^{j_1j_2}(n_1,n_2,t)  \nonumber \\ 
& &=\sum\limits_{j_1{'}j_2{'}} M_h^{j_1{'}j_2{'}}(n_1,n_2,0) M_{j_1{'}}^{j_1}(n_1,t) M_{j_2{'}}^{j_2}(n_2,t)\nonumber \\
& &+M_{h({\rm QCD})}^{j_1j_2}(n_1,n_2,t),
\end{eqnarray}
where
\begin{eqnarray}
\label{1.36}
& &M_{h({\rm QCD})}^{j_1j_2}(n_1,n_2,t) \nonumber\\
& &=\sum\limits_{i} 
M_h^i (n_1+n_2,0) M_{i}^{j_1j_2}(n_1,n_2,t) 
\end{eqnarray} 
are the particular solutions of the complete equations with zero initial conditions at the hadron level, and
\begin{eqnarray}
\label{1.36q}
M_{i}^{j_1j_2}(n_1,n_2,t) = \sum\limits_{jj_1{'}j_2{'}} \int\limits_0^t dt{'}
M_i^j (n_1+n_2,t{'})\nonumber \\
\times P_{j \to j_1{'}j_2{'}}(n_1,n_2) 
 M_{j_1{'}}^{j_1}(n_1,t-t{'}) M_{j_2{'}}^{j_2}(n_2,t-t{'})
\end{eqnarray}
are the particular solutions of the complete equations with zero initial conditions at the parton level. The first term in the expression~(\ref{1.37}) represents the solutions of homogeneous evolution equations with the given initial conditions
$M_h^{j_1{'}j_2{'}}(n_1,n_2,0)$. These unknown nonperturbative two-parton initial conditions just reckon for the unsolved confinement problem. If one assumes that there is the approximate factorization of these initial conditions in the moment representation
\begin{eqnarray}
\label{fact in}
M_h^{j_1{'}j_2{'}}(n_1,n_2,0)\simeq M_h^{j_1{'}}(n_1,0) M_h^{j_2{'}}(n_2,0),
\end{eqnarray}
then the solutions of the homogeneous evolution equations will be approximately factorized in $x$ representation as well.

Now we consider the initial condition effects in the asymptotic behavior ($t \to \infty$) of the dPDFs. The equations (\ref{1.37}), (\ref{1.36}) and (\ref{1.36q}) show that the initial conditions are related to the solutions with different dependence on evolution variable $t$. In order to better understand the character of this dependence, at first we use a toy model with one type of partons (for instance, QCD theory with gluons only, or six-dimensional $\phi^3$ theory~\cite{konishi0,konishi}). In this case the dPDFs become simpler and look like
\begin{eqnarray}
\label{1.38}
& &M_h^{11}(n_1,n_2,t) = M_h^{11}(n_1,n_2,0) \exp\{[P(n_1) + P(n_2)]t\}  \nonumber \\
& &+\frac{P(n_1,n_2)M_h^1(n_1+n_2,0)}{P(n_1+n_2)-P(n_1)-P(n_2)}
\{\exp[P(n_1+n_2)t]  \nonumber \\
& &- \exp[(P(n_1)+P(n_2))t]\}. 
\end{eqnarray}
Thus, for $t$ large enough, we have two different asymptotic regimes depending on the relation between the kernels $P(n_1+n_2)$ and $P(n_1) + P(n_2)$:

1. If $P(n_1+n_2) < P(n_1) + P(n_2)$, then
\begin{eqnarray}
\label{1.40}
& & M_h^{11}(n_1,n_2,t)|_{t \to \infty} = \Bigg[M_h^{11}(n_1,n_2,0) +\nonumber\\
& &\frac{P(n_1,n_2)M_h^1(n_1+n_2,0)}{P(n_1)+P(n_2)-P(n_1+n_2)}\Bigg] 
\nonumber \\
& & \times \exp\{[P(n_1) + P(n_2)]t\}. 
\end{eqnarray}
2. If $P(n_1+n_2) > P(n_1) + P(n_2)$, then
\begin{eqnarray}
\label{1.42}
 M_h^{11}(n_1,n_2,t)|_{t \to \infty} & = &\frac{P(n_1,n_2) 
M_h^1(n_1+n_2,0)}{P(n_1+ n_2) - P(n_1) - P(n_2)} 
\nonumber \\
& & \times \exp[P(n_1+n_2)t].
\end{eqnarray}
For the second regime, the asymptotic behavior of dPDFs {\it does not dependent} on the initial correlation conditions $M_h^{11}(n_1,n_2,0)$ at all, and is specified by the correlations perturbatively calculated.

In the toy $\phi^3_6$-model~\cite{konishi0,konishi}
\begin{equation}
\label{1.44}
P(n) = \frac{1}{(n+2)(n+3)}-\frac{1}{12},
\end{equation}
and both asymptotic regimes are realized. However, if we are interested in the dPDFs in the region of finite $x_1$ and $x_2$, we conclude that the dPDFs "forget" the initial correlation conditions because their asymptotic behavior in this region is determined by large moments $n_1$ and $n_2$, where $P(n_1+n_2) > P(n_1) + P(n_2)$.

The presence of several parton types does not essentially complicate the analysis of the asymptotic behavior of the dPDFs. Indeed, in this case one has to express
single parton distributions via the eigenfunctions of corresponding DGLAP equations (see, for instance, Refs.~\cite{dokshitzer, konishi, dok2}), put them into Eqs. 
(\ref{1.37}), (\ref{1.36}) and (\ref{1.36q}) and take the leading contributions into consideration only. As a result, the relation between maximum
eigenvalues $\Lambda(n_1+n_2)$ and $\Lambda(n_1)+\Lambda(n_2)$ will determine the asymptotic behavior regime of the dPDFs:

1. If $\Lambda(n_1+n_2) < \Lambda(n_1) + \Lambda(n_2)$, then the dPDFs are dependent on the initial correlation conditions $M_h^{j_1 j_2}(n_1,n_2,0)$.

2. If $\Lambda(n_1+n_2) > \Lambda(n_1) + \Lambda(n_2)$, then the dPDFs are independent of the initial correlation conditions $M_h^{j_1 j_2}(n_1,n_2,0)$.

The eigenvalues and the eigenfunctions for the single distributions in QCD have been thoroughly studied~\cite{dokshitzer,konishi,dok2}. The results of these studies show that in QCD as well as in the case of a model example both asymptotic regimes are realized. Therefore, one needs to know the initial correlation conditions (which, generally speaking, are arbitrary and should be extracted from the experiment) to determine even the asymptotic behavior of the dPDFs. However, we come again to the relation
$$ \Lambda(n_1+n_2) > \Lambda(n_1) + \Lambda(n_2)$$
for large moments $n_1$ and $n_2$ that determines the dPDFs in the region of not parametrically small $x_1$ and $x_2$, because $\Lambda(n) \sim -\ln(n), n \gg 1$ (the $n$-dependence of the eigenvalues $\Lambda(n)$ can be found in detail, for instance, in Refs.~\cite{dokshitzer, konishi,dok2}). 

\section{\label{sec4}Conclusions}

Using the explicit form of the solutions of evolution equations in the Mellin representation we conclude that the dPDFs "forget" the initial correlation conditions (unknown {\it a priori}) at not parametrically small longitudinal momentum fractions, and the correlations perturbatively calculated survive only in the limit of large enough hard scales. Such a dominance is the mathematical consequence of the relation between the maximum eigenvalues $\Lambda(n)$ in the moment representation, ~$\Lambda(n_1+n_2)~>~\Lambda(n_1)+\Lambda(n_2)$, in QCD at large $n_1$ and $n_2$ (finite $x_1$ and $x_2$). It is independent of the strength of the initial correlation conditions.

The asymptotic behavior analysis indicates a tendency only, but tells nothing definite about the values of $x_1,x_2$ and $t(Q^2)$ where the correlations become significant and the asymptotic behavior appears to be a good approximation to the reality. On the other hand, the numerical estimates~\cite{Gaunt:2009re,Diehl:2014vaa} testify in favor of conclusion that the perturbative correlation effects are significant in the kinematical region accessible in experimental measurements at energies of Tevatron and LHC. 

It is of interest also to know the magnitude of these additional perturbative contributions to the real processes. For instance, the term with correlations may entail~\cite{Cattaruzza:2005nu} a correction of the order of 40$\%$ to the multiple $W$ production cross sections in $pp$ collisions at $\sqrt{s}=1$ TeV, and of the order of 20 $\%$ at $\sqrt{s}=14$ TeV. In the case of $b\bar b$ pairs production the correction terms are of the order of 10-15$\%$ at 1 TeV and of the order of 5$\%$ at 14 TeV. Similar values were obtained in the frameworks of the Monte-Carlo generator PYTHIA~\cite{pythia,pythia8} for the so-called joined interactions~\cite{sjostrand3} (in the authors' terminology), which means averaging at roughly one joining per 15 events for the minimum-bias sample and one per 7 events for the underlying event (with $p_{\perp \rm hard} >$ 100 GeV) sample for $p\bar p$ collisions at $\sqrt{s}=1.96$ TeV. Although no algorithm implementing the full kinematics for such joined interactions has been constructed yet, the Monte-Carlo generator PYTHIA describes successfully almost all experimentally available data on DPS, since it realistically takes into account a number of essential correlations~\cite{sjostrand2} in flavor, color, longitudinal and transverse momentum, thereby compensating, in a sense, the effects due to additional (omitted) contributions.   

Unlike naively accepted expectations, the QCD dynamical correlations imply the dependence~\cite{Snigirev:2010tk,Ryskin:2011kk,flesburg,Blok:2013bpa} of the experimentally extracted effective cross section of DPS on the resolution scale. The measurements covering a larger range of the resolution scale variation might reveal the evolution effects more distinctly in accordance with the numerical estimations in Ref.~\cite{Blok:2013bpa} and the asymptotic QCD behavior considered above.

\begin{acknowledgments}
Discussions with K.O.~Bugaev, A.I.~Demyanov, S.V.~Molodtsov, O.V.~Teryaev and N.P.~Zotov are gratefully acknowledged. This work is partly supported by Russian Foundation for Basic Research Grant No. 13-02-01005 and National Academy of Sciences of Ukraine, Grant TSO-1-2/2014.
\end{acknowledgments}


\end{document}